\documentclass[aps,prb,preprint,aps]{revtex4-2} 

\usepackage{etoolbox}
\makeatletter
\newcounter{mycitecount}
\newcommand{\mycite}[1]{%
  [%
  \setcounter{mycitecount}{0}%
  \forcsvlist{\myciteaux}{#1}%
  ]%
}
\newcommand{\myciteaux}[1]{%
  \ifnum\value{mycitecount}>0,\fi
  \@ifundefined{b@#1}{??}{\csname b@#1\endcsname}%
  \stepcounter{mycitecount}%
}
\makeatother

\usepackage{amsmath}  
\usepackage{amsfonts} 
\usepackage{graphicx} 
\usepackage{float}
\usepackage{xcolor}
\usepackage{soul}

\usepackage[dvipsnames]{xcolor} 

\begin{document}


\title{Blowing in a drinking straw: introducing Quantum Physics.}

\author{Lorenzo Galante}
\email{lorenzo.galante@polito.it} 
\affiliation{Politecnico di Torino, TLlab, 10129, Torino, Italy}


\date{\today}

\begin{abstract}
We present an educational approach aimed at introducing the fundamental concepts of quantum mechanics (QM). By exploiting the formal analogy between sound wave behavior in an acoustic pipe (a drinking straw) and the quantum infinite square potential well, we provide an intuitive framework that explains the origin of energy quantization in bound quantum systems (such as atoms), introduces the concept of wavefunction, and lays the groundwork for discussing the Heisenberg uncertainty relations. The proposed method enables low-cost experimental activities that actively involve students, facilitating a meaningful connection between theoretical principles and empirical observations. Moreover, it encourages critical reflection on the Copenhagen interpretation of the wavefunction, promoting a deeper conceptual understanding.\\
\textbf{Keywords:} Quantum Physics Education, Analogy, Uncertainty, Wave Function, Hydrogen Energy Levels.
\end{abstract}
\maketitle 

\section{Introduction}\label{sec:I}
Engaging students in a hands-on exploration of the quantized frequency spectrum of a simple drinking straw offers an outstanding entry point for introducing fundamental concepts of QM \mycite{bib4,bib2}. Each group receives an identical straw and uses a smartphone app to analyze the frequencies produced; the collaboratively collected data, entered into a shared spreadsheet, yield reliable estimates of the system's first resonant frequencies. By discussing the behavior of air in the pipe that is open at both ends, the class can link the quantized spectrum to the more general problem of a wave confined within a finite spatial region, thus achieving a first educational milestone: quantization occurs whenever a wave is trapped in a limited space. To reinforce this idea, students can investigate how the resonant frequencies change as the straw length increases, ultimately approaching the limit of an infinitely long straw. They observe that, when the confining boundaries disappear, the discrete levels crowd together until quantization effectively vanishes.

If the teacher deems it appropriate, students may also tackle the formal mathematical treatment of the problem, an elegant demonstration of how the calculus tools they have learned are essential for modeling and solving the acoustical pipe.

The connection with quantum systems such as atoms descends naturally, highlighting the educational potential of acoustic-quantum analogies in supporting conceptual understanding \mycite{bib3, bib8, bib7}. The analogy with the pipe opened at both ends rests on two key facts:
\begin{enumerate}
    \item both the acoustic and the quantum case are described in terms of waves \mycite{bib12}.
    \item the atom is spatially confined by a potential well just as the pressure wave in the straw is confined between the two open ends.
   
\end{enumerate}

Moreover, the D'Alembert equation governing the time and space-dependent pressure inside the straw is formally identical to the stationary Schr\"odinger equation for a particle in an infinite square potential well, a pedagogical idealization of the hydrogen atom potential.

The pressure mode shapes associated with each frequency mirror the wavefunctions of the energy levels of the quantum problem, providing an excellent opportunity to draw a parallel as a starting point to introduce the Copenhagen interpretation.

In the following sections, we present and discuss a possible sequence of activities based on this approach. The sequence has been implemented in three separate courses (five hours each), involving roughly 100 students overall.

\section{Resonant frequencies of a drinking straw.}\label{sec:II}
\begin{figure}[H]
  \centering
    \includegraphics[scale=0.35]{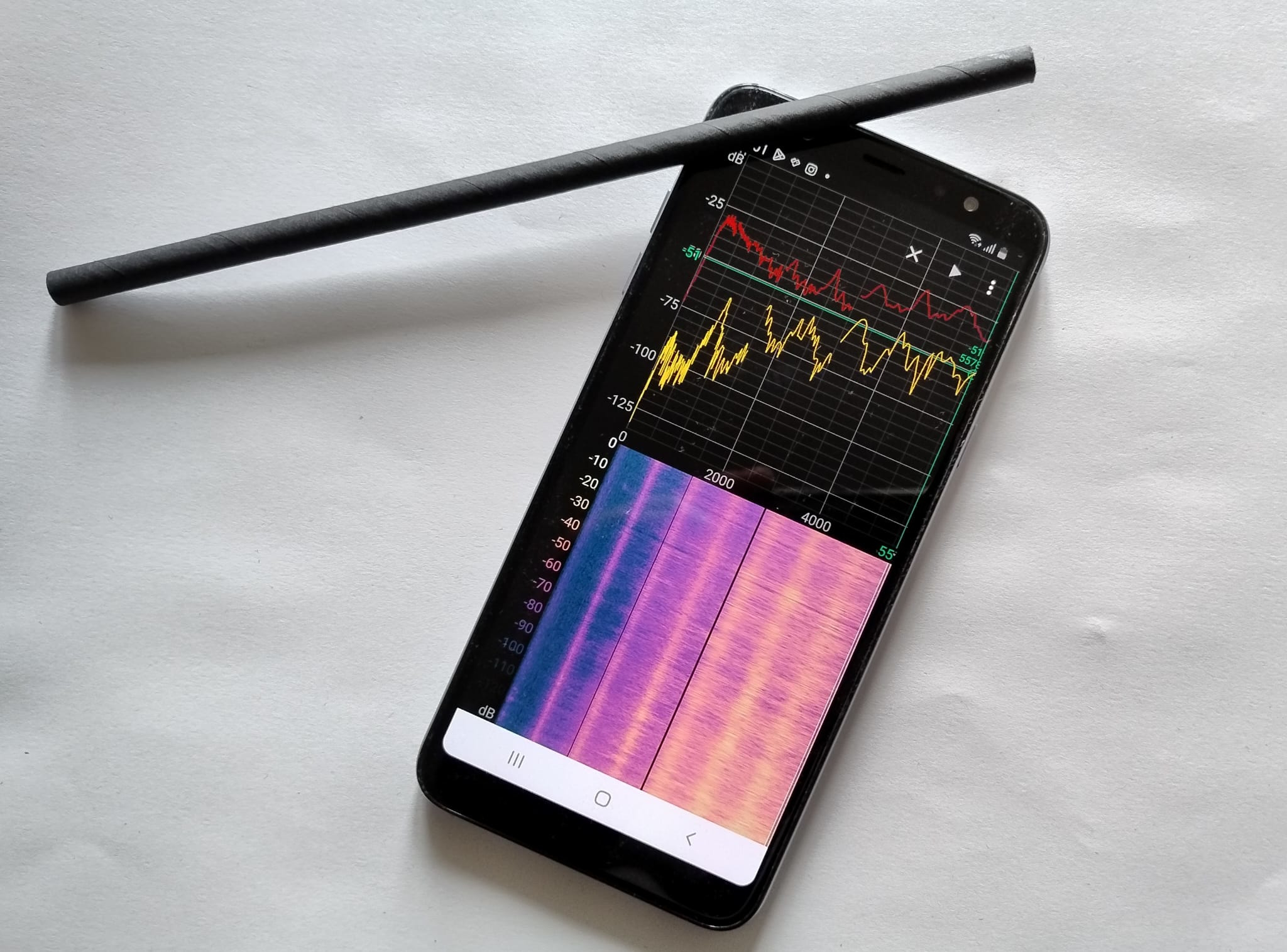}
  \caption{The experimental apparatus used to engage students in the measurements of the resonant frequencies. A simple drinking straw and a free app as a spectrum analyzer (Spectroid, for Android OS; Spectrumview, for IOS).}
  \label{img:dev}
\end{figure}

\begin{figure}[H]
  \centering
    \includegraphics[scale=0.2]{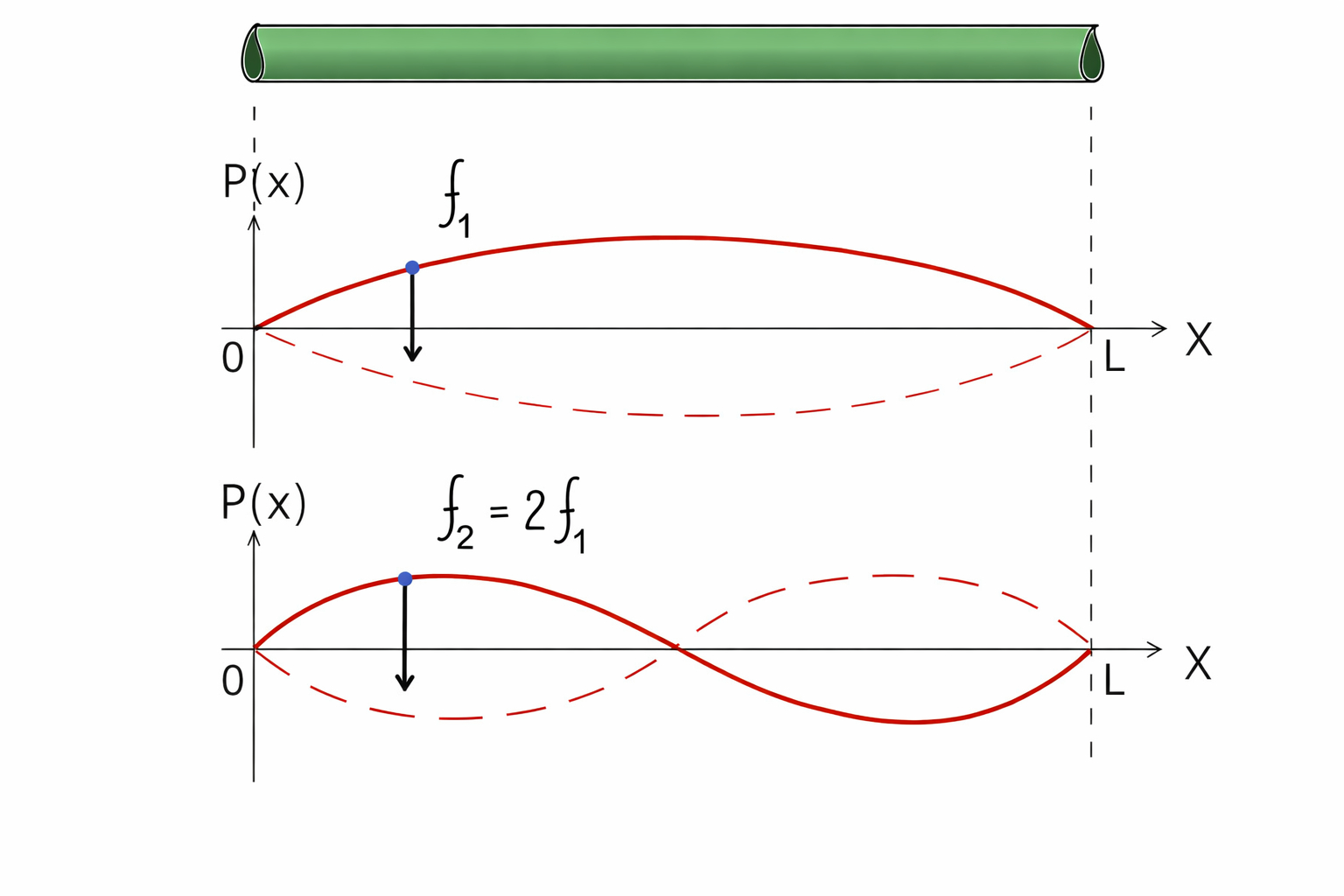}
  \caption{The first two pressure modes of the straw. The first one oscillates at frequency $f_1=\frac{c}{\lambda_1}$ related to the wavelength $\lambda_1=2L$.}
  \label{img:mod}
\end{figure}

The first classroom activity explores the resonant spectrum of an air column in a pipe opened at both ends. Each student receives an identical straw, L = 19.9 cm, and installs a free spectrum-analysis app on their smartphone \mycite{bib10} (Fig. \ref{img:dev}). 
A brief introduction reminds students that standing waves inside a pipe depend on two independent variables: position $x$ along the axis and time $t$. At any fixed $x$, the acoustic pressure oscillates harmonically at frequency $f$. Spatial and temporal periodicities are linked through
\begin{equation}
    \lambda=c\,T=\frac{c}{f}\,,
    \label{eq:lam1}
\end{equation}
where $\lambda$ is the wavelength, $c$ the speed of sound and $T$ the period.
Because the acoustic pressure at both open ends must equal ambient pressure (set to zero), standing waves arise only when an integer number of half-wavelengths fits into the pipe (Fig. \ref{img:mod}):
\begin{equation}
    n\frac{\lambda}{2}=L\,,\quad f_n=n\cdot\frac{c}{2L}\,,\quad (n=1, 2, ...)\,.
    \label{eq:lam2}
\end{equation}
Thus the straw exhibits an equally spaced (quantized) frequency spectrum $f_n$.

\begin{figure}[h]
  \centering
    \includegraphics[scale=0.5]{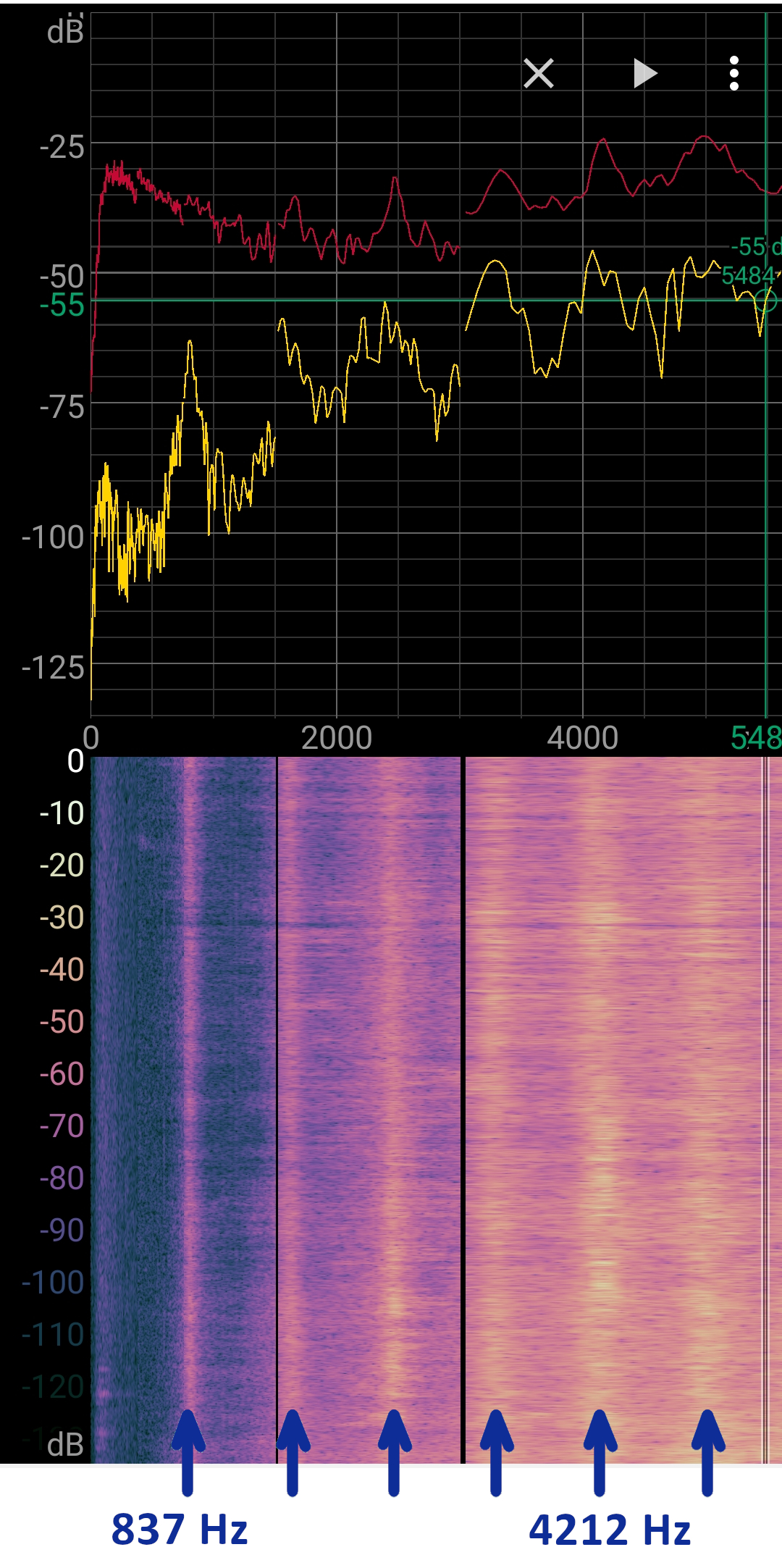}
  \caption{The spectrogram clearly shows the evenly spaced frequencies of the straw. The value of the frequency corresponding to each vertical line is given on the   axis just above the spectrogram. Measurements are carried out by tapping on the screen and positioning a cursor on each vertical line. }
  \label{img:app1}
\end{figure}

\begin{figure}[H]
  \centering
    \includegraphics[scale=0.5]{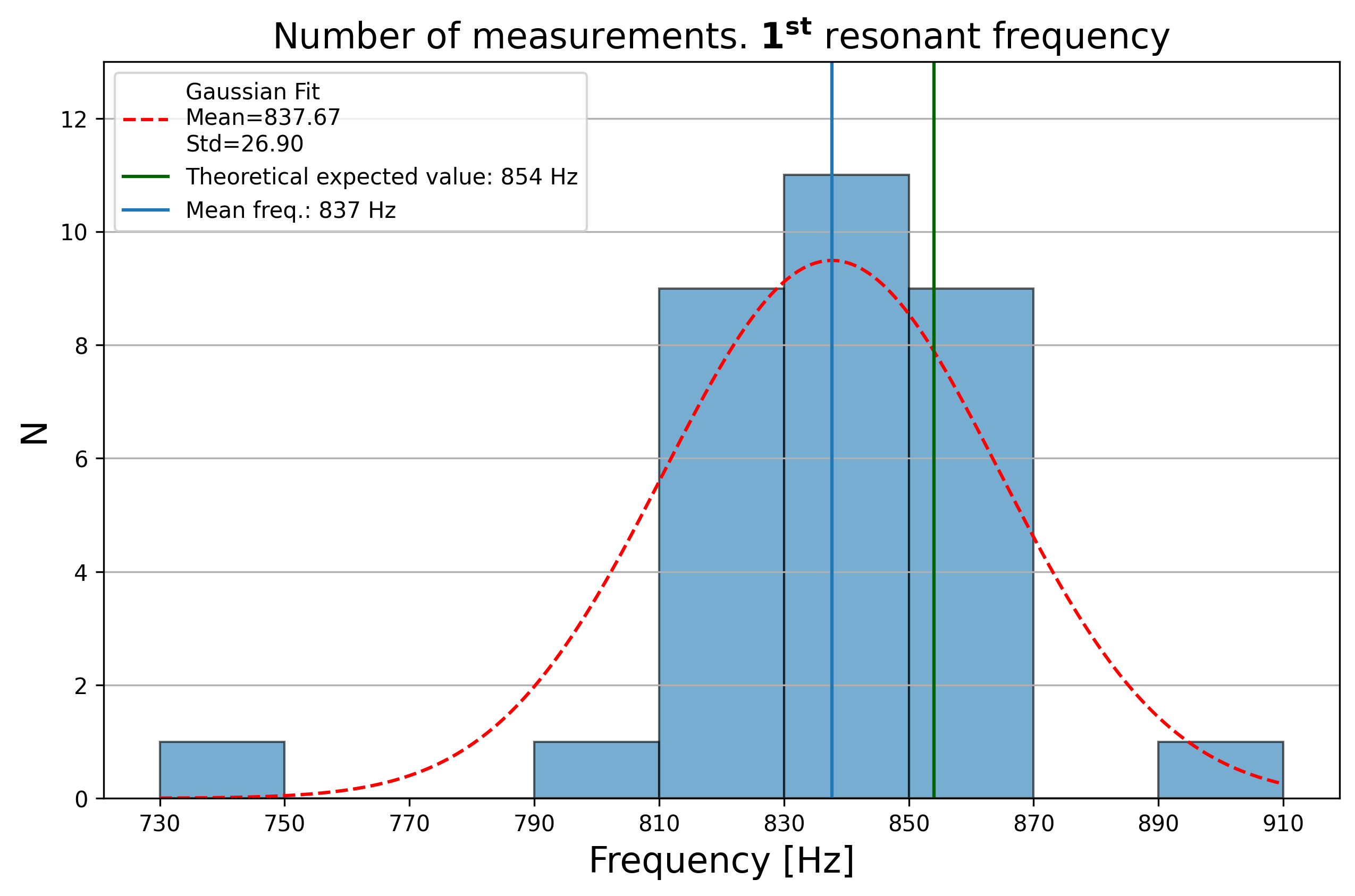}
  \caption{Statistical distribution of the results for the first resonant frequency.}
  \label{img:stat}
\end{figure}

Students are asked to measure the resonant frequencies of the first five modes of the straw (Fig. \ref{img:app1}) and to share their results on a collaborative spreadsheet where the mean values of their measurements are evaluated and compared to the theoretical expected values. In order to excite all the resonant modes at the same time, students are told to blow across one open end, while the phone's microphone, placed near the opposite end, records the sound (the best results are obtained when blowing at an angle of roughly $90^{\circ}$ relative to the straw axis). The experiment works even in the presence of classroom background noise. However, once the measurement procedure has been introduced, the teacher can also assign the experimental task as homework. The experimental uncertainty is reported in Table \ref{tab:fr}.
Table \ref{tab:fr} lists the average values of the first five resonant frequencies measured by the 32 student teams in the three courses and compares them with the corresponding theoretical predictions (computed with $c=340\,\,ms^{-1}$). To convey the overall quality of the experimental campaign, Fig. \ref{img:stat} depicts the full statistical distribution for the first resonant frequency $f_1$. In the histogram, the sample mean (blue vertical line), its standard deviation, a gaussian fit to the data, and the the theoretical value (green vertical line) are indicated for easy comparison. 
It's worth noting that the theoretical frequencies are slightly higher than the experimental averages. This offset is expected and easy to explain. The simplified  model we apply ignores the classical end correction associated with the finite radius of the straw’s two open ends \mycite{bib13}. Accounting for this effect makes the effective acoustic length
\begin{equation}
L_{eff}=L+2\cdot( 0.6\,\,r)
    \label{eq:endeff}
\end{equation}
where $r$ is the straw’s inner radius. Because a longer column lowers the resonant frequencies, the omission of $2\cdot( 0.6\,r)$ in $L_{eff}$ causes every theoretical prediction to be slightly above the measured values. Whether to present this refinement in class is a pedagogical decision: the core learning objectives do not depend on it. One attractive option is to let students discover the systematic shift themselves, visible across all harmonics, and challenge them to formulate a scientific explanation.

\begin{table}[h!]
\centering
\caption{First five resonant frequencies.}
\begin{ruledtabular}
\begin{tabular}{ c c c c c}
\multicolumn{5}{c}{\textbf{Measured values of the resonant frequencies [Hz]}}   \\ \hline	
 \phantom{000}837 & \phantom{0000}1654 & \phantom{0000}2509 & \phantom{00000}3331 & 4212 \\ \hline
\multicolumn{5}{c}{\textbf{Theoretical values [Hz]} }  \\ \hline  
\phantom{000}854 & \phantom{0000}1709 & \phantom{0000}2563 & \phantom{00000}3417 & 4271 \\ \hline
\multicolumn{5}{c}{\textbf{Percentage Variation $(f_{Theor.}-f_{Meas.})/f_{Theor.}$}}   \\ \hline  
\phantom{000}0.02 & \phantom{0000}0.03 & \phantom{0000}0.02 & \phantom{00000}0.03 & 0.01
\end{tabular}
\end{ruledtabular}
\label{tab:fr}
\end{table}

\section{A step deeper in the physics of a straw}\label{sec:III}
\begin{figure}[H]
  \centering
    \includegraphics[scale=1.8]{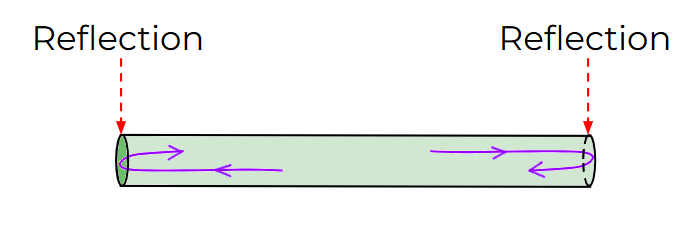}
  \caption{In this model, the pressure wave reflects at both ends, so it is trapped in a finite space.}
  \label{img:refl}
\end{figure}

In the first activity students needed only a qualitative picture of standing waves to locate and measure the resonant peaks of a drinking straw. At this point, however, the data they collected can serve as a springboard for a more thorough physical analysis.
We begin by revisiting what happens at each open end of the pipe. When an acoustic disturbance traveling inside the straw reaches an abrupt change in cross-sectional area, the mismatch in acoustic impedance forces the wave to reflect back into the tube \mycite{bib13}. Because reflections occur at both ends, the pressure wave is effectively trapped and the air column becomes a finite resonator (Fig. \ref{img:refl}).
With this picture in place, the two ingredients that generate the quantized spectrum measured earlier are now unmistakable:
\begin{enumerate}
    \item Wave behavior: the acoustic pressure obeys a wave equation.
    \item Spatial confinement: the wave is restricted to a finite region.
\end{enumerate}

Whenever these two conditions coexist, some physical quantity (frequency, in our case) becomes quantized. 

\subsection{An extension activity}
\begin{figure}[H]
  \centering
    \includegraphics[scale=0.4]{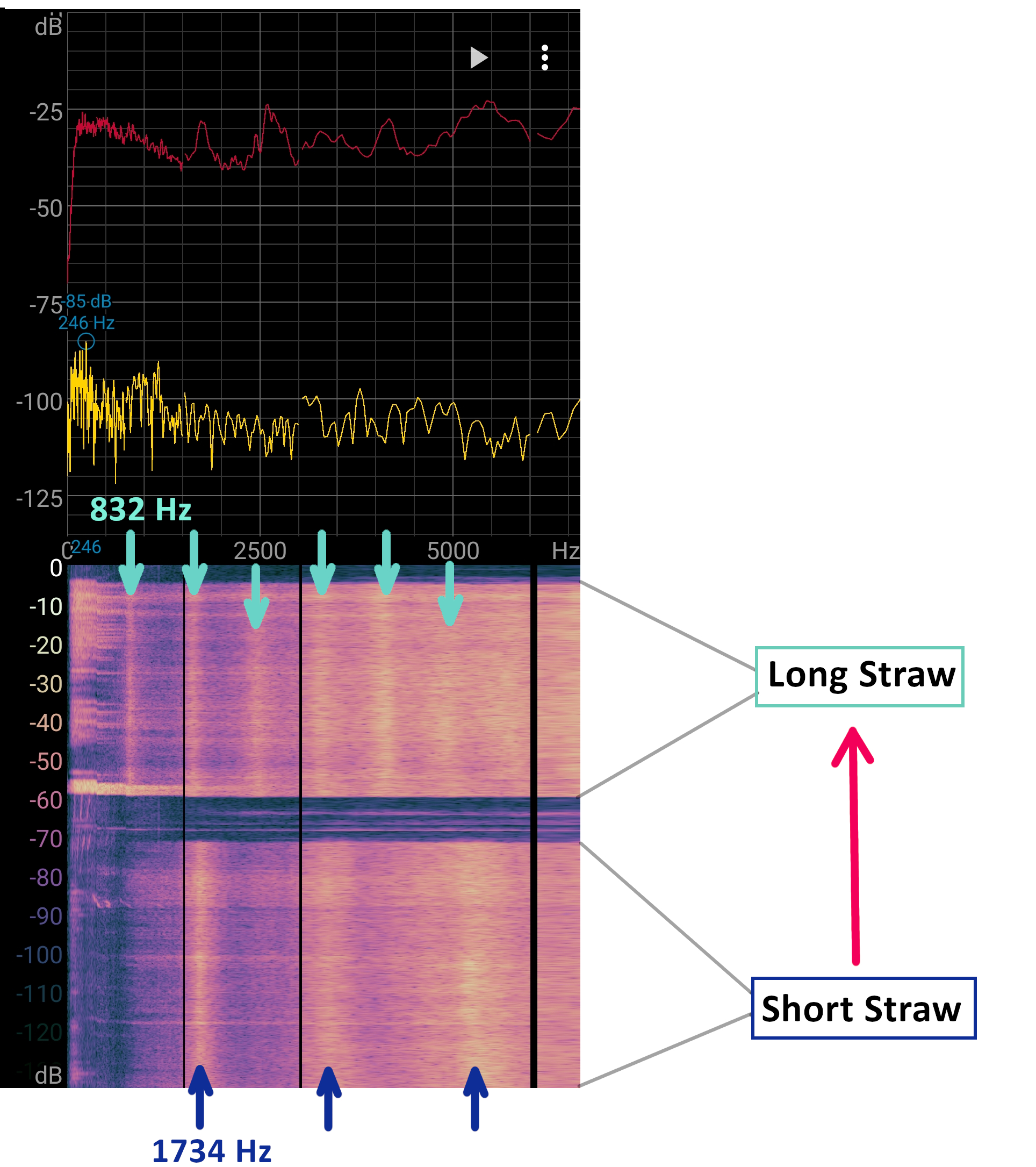}
  \caption{The spectrum of a short straw (dark blue arrows) and of a longer straw (light blue arrows) is compared and measured. Moving from a more confined system to a less confined one, the frequency levels become more and more close to each other. In a limiting process, students might grasp that for a not confined system ($L\rightarrow\infty$) quantization should vanish.}
  \label{img:lim}
\end{figure}

To make the link between confinement and level spacing tangible, students can repeat the measurement with straws of increasing length. They quickly discover that the gap between successive resonances shrinks as $L$ grows; in the limiting case $L\rightarrow\infty$  the spectrum becomes so densely packed that quantization disappears and a continuum emerges (Fig. \ref{img:lim}).

The groundwork has been laid for extending the discussion to bound quantum systems, such as the atom.

\section{The quantum analogy: from the straw to the atom}\label{sec:IV}
\begin{figure}[H]
  \centering
    \includegraphics[scale=0.9]{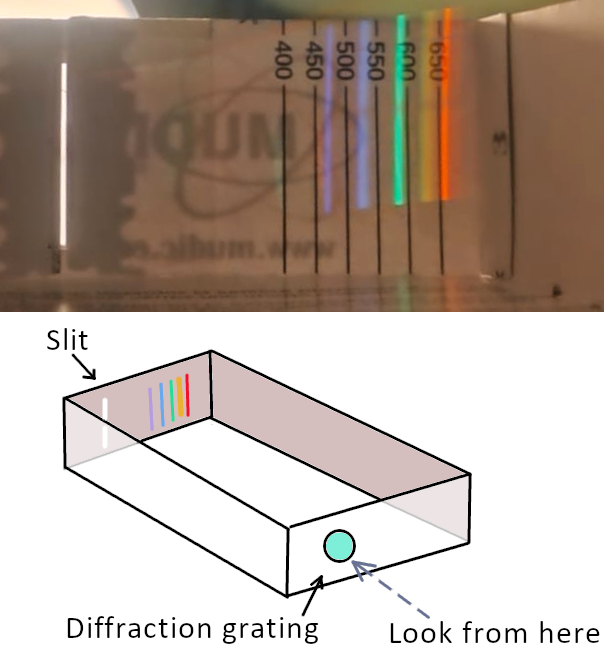}
  \caption{Top: the spectrum of the light source as was observed in the classroom. Bottom: the self-built card-stock spectrometer. Looking through the circular window, where the diffraction grating is located, the spectrum appears inside the box next to the slit. }
  \label{img:spec}
\end{figure}

In the previous section, we observed that quantization arises whenever a system is governed by wave physics and is spatially confined. Precisely these same factors appear in QM. Indeed, an atom is a confined system. Consider the simplest example of the Hydrogen atom, in which an electron is bound to a proton \mycite{bib5}. The fundamental equation describing this quantum system is the Schr\"odinger equation, a wave equation involving the variable $\Psi$, the wavefunction. Therefore, it should not come as a surprise that quantization emerges also in atomic systems.
At this stage, observing the discrete emission spectrum from atoms becomes a valuable teaching experience. Such experiments can be easily conducted in the classroom, even using low-cost equipment or self-built apparatus. In our courses we used a self-built spectrometer and a common fluorescent lamp (Fig. \ref{img:spec}). The direct observation of the discrete spectrum of light coming from atoms closes the first educational loop. Even at an introductory level, the primary objective can be considered achieved: students should have grasped the underlying reasons for the quantization of energy levels in atoms. The validity of these reasons is supported and illustrated through the acoustic analogy and the straw experiment. The fact that atoms meet the same essential conditions leading to quantization is justified both by the well-established understanding that atoms are spatially confined systems, and by the direct observation of the discrete spectrum of light they emit. In every case, the straw experiment supplies a concrete mental model that makes the abstract idea of energy levels quantization less intimidating.
\section{From concepts to equations: a deeper insight into the analogy}\label{sec:V}
So far, the analogy between acoustic systems and quantum systems has served as a powerful conceptual bridge, offering students an intuitive grasp of energy quantization \mycite{bib9}. However, beneath the surface of this conceptual parallel lies a deeper, mathematical structure that further strengthens the analogy and reveals its true explanatory power.
In this section, we transition from qualitative reasoning to quantitative analysis. By examining the wave equations governing both sound in a straw and the problem of a quantum systems confined in space (like atoms), we uncover a striking structural similarity. Both systems are described by wave functions constrained by boundary conditions within a finite domain, leading naturally to discrete solutions, or quantized modes. This mathematical parallel not only reinforces the conceptual analogy, but also highlights how physical intuition can be elegantly captured and confirmed through formalism. It is here that students begin to see how mathematical tools are not mere abstractions, but essential instruments for uncovering and articulating the laws of nature. It is therefore essential to link these concepts explicitly to what students learn in calculus, since the analogy is grounded in fundamental mathematical tools such as derivatives and differential equations.

\subsection{Modeling the Hydrogen atom with an infinite square well}

\begin{figure}[H]
  \centering
    \includegraphics[scale=0.7]{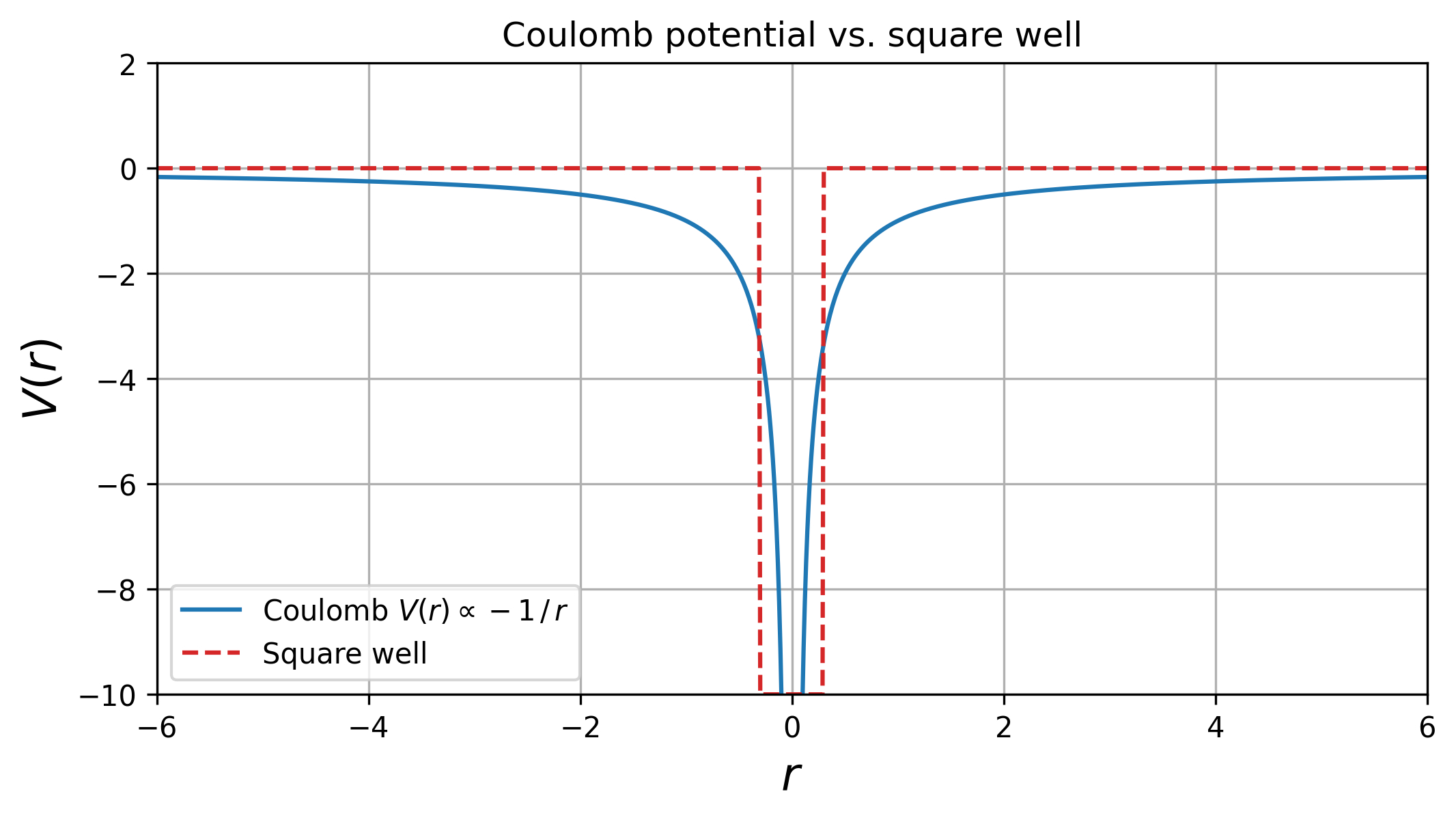}
  \caption{The Coulombian potential approximated by an infinite square well (arbitrary units).}
  \label{img:coul}
\end{figure}

The central idea is that, in a simplified educational model, the Coulomb potential of the simplest atomic system (the Hydrogen atom) can be approximated by an infinite square potential well (see Fig. \ref{img:coul}). Starting from this assumption, we can introduce several cornerstone notions of confined quantum systems: the quantization of energy levels, the concept of the wavefunction, and even a first glimpse of the Heisenberg uncertainty relations. This idealization is especially valuable in teaching for two reasons:
\begin{enumerate}
    \item Mathematical simplicity. From a mathematical point of view, the quantum problem of a particle confined within an infinite square well is among the most straightforward and analytically solvable cases (potentially accessible even to students in the final year of secondary school),
    \item Formal analogy with classical waves. The time-independent Schr\"odinger equation for this system is mathematically identical to the wave equation governing air‑pressure oscillations inside a straw.
\end{enumerate}
\subsection{Formal identity with the straw wave equation}
We can briefly explore the formal analogy by noticing that, for a general potential barrier, the one dimensional time-independent Schr\"odinger equation is:
\begin{equation}
-\frac{\hbar^2}{2m}\frac{d^2\Psi}{d x^2} +V(x)\,\Psi=E\,\Psi\,,
\label{eq:schr}
\end{equation}
where $m$ and $E$ are the mass and energy of the particle, $V(x)$ the potential energy as a function of the position $x$ and $\hbar$ is the Planck constant divided by $2\pi$. The equation can also be written in the form of an eigenvalue problem:
\begin{equation}
H\Psi=E\,\Psi\,,
\label{eq:Hamilt}
\end{equation}
where $H$ is the Hamiltonian operator
\begin{equation}
H=-\frac{\hbar^2}{2m}\frac{d^2}{d x^2} +V(x)\,.
\label{eq:Hamilt1}
\end{equation}

\begin{figure}[H]
  \centering
    \includegraphics[scale=0.3]{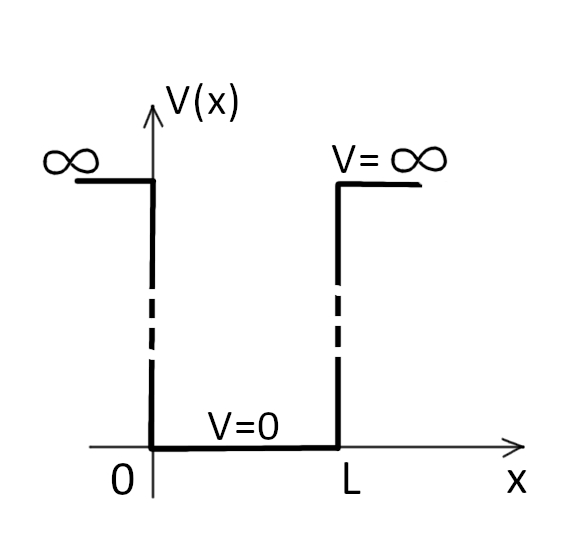}
  \caption{The shape of an infinite square well.}
  \label{img:rect}
\end{figure}

In the case of an infinite square potential well of width $L$ (Fig. \ref{img:rect}), we set $V(x)=0$ for $0<x<L$ and $V=\infty$ elsewhere. Inside the well, Eq. (\ref{eq:schr}) reduces to:
\begin{equation}
-\frac{\hbar^2}{2m}\frac{d^2\Psi}{d x^2} =E\,\Psi\,,
\label{eq:schr1}
\end{equation}
and can be rearranged in the following form:
\begin{equation}
\frac{d^2\Psi}{d x^2} =-k^2\,\Psi\,,\quad \text{where}\quad k^2=\frac{2mE}{\hbar^2}\,.
\label{eq:schr2}
\end{equation}

At this point the relation with the acoustics of a straw is evident, since the wave equation describing air pressure inside a straw is the D'Alembert equation:  
\begin{equation}
\frac{\partial^2p}{\partial x^2}-\frac{1}{c^2}\frac{\partial^2 p}{\partial t^2}=0\,,
    \label{eq:dalemb1}
\end{equation}
where $p = p(x, t)$ represents air pressure as a function of position and time and $c$ is the speed of sound. By searching for harmonic solutions of the form $p(x, t) = p(x)\cdot \cos(\omega t)$, the equation simplifies to: 
\begin{equation}
\frac{d^2p}{d x^2}= -k^2\,p(x)\,,\quad \text{with}\quad k^2=\frac{\omega^2}{c^2}\,,
    \label{eq:dalemb2}
\end{equation}
formally identical to Eq. (\ref{eq:schr2}). 

Due to the infinity of the potential well, the particle with mass $m$ is trapped in the region $L$ exactly in the same way the pressure wave is within the straw. By solving  Eqs. (\ref{eq:schr2}) and (\ref{eq:dalemb2}), for the quantum or the acoustical problem, we obtain how the wavefunction $\Psi$ or the pressure $p$ are distributed along $x$ in the domain $x\in[0,\,L]$. 

A straightforward solution to these equations is: 
\begin{equation}
p(x)=A\,sin(k\,x)\,\quad \text{and}\quad\Psi(x)=A\,sin(k\,x)\,.
    \label{eq:sol_dalem}
\end{equation}
Note that the solution chosen here is not the most general, which would typically include a superposition of both sine and cosine terms. Nevertheless, this simpler approach effectively captures the key physics underlying the problem. Importantly, the choice of this simpler form does not alter the solutions.

\subsection{Boundary Conditions and Quantized Solutions}
As anticipated, the formal identity of the two cases allows us to extend the well known results of the pipe to the quantum system trapped in the potential well.
Boundary conditions are basically the same: the pressure at the end of the pipes must equate the external atmospheric pressure that we set to $0$, the wavefunction at the boundaries and in the external region has to be $0$, since no particle can escape an infinitely deep potential well. By imposing these conditions:
\begin{equation}
p(0)=p(L)=0\quad\text{and}\quad \Psi(0)=\Psi(L)=0,
    \label{eq:booundary}
\end{equation}
quantization of the number $k$ arises naturally and, given that $k=2\pi/\lambda$, also the wavelength results to be quantized 
\begin{equation}
k_n=\frac{\pi}{L}\cdot n\quad,\quad \lambda_n=\frac{2L}{n}\,,\quad (n=1, 2, ...)\,.
    \label{eq:k_quant}
\end{equation}
The solutions of the Schr\"odinger equation have the same shape of the pressure modes of a pipe (Fig. \ref{img:mod}). Note that in this context, the solutions are the so called eigenfunctions of the Hamiltonian operator for the infinite square well, each one associated to a different value, eigenvalue, of $k_n$ and consequently of the energy. More generally, a wavefunction represents the quantum  state of a system, while eigenfunctions correspond to particular pure-state wavefunctions associated with a definite eigenvalue of a physical observable. 

\begin{figure}[H]
  \centering
    \includegraphics[scale=0.3]{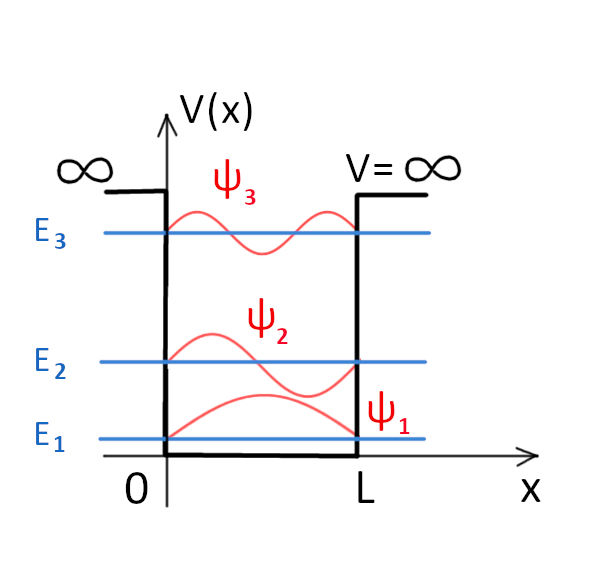}
  \caption{The quantized spectrum of the square box problem and the corresponding eigenfunctions $\Psi_n(x)$.}
  \label{img:boxlev}
\end{figure}

We have a different eigenfunction for each value of $k_n$:
\begin{equation}
\Psi_n(x)=A\, sin(\frac{\pi}{L} n\,x)
    \label{eq:psi_quant}
\end{equation}
From Eqs. (\ref{eq:schr2}) and (\ref{eq:dalemb2}) we know that $k$ is related to energy $E$, for the quantum problem, and to the angular frequency $\omega$, for the pipe case. As a result, we immediately understand that the quantization of $k$ implies energy quantization for the particle and frequency quantization for the sound in a pipe. Each eigenfunction $\Psi_n$ corresponds to a different energy level $E_n$ of a discrete spectrum (Fig. \ref{img:boxlev}) and each pressure mode $p_n$ corresponds to a different oscillation frequency $f_n$:
\begin{equation}
    E_n=\frac{\hbar^2k_n^2}{2m}=\frac{h^2}{8mL^2}\cdot n^2\,,\quad f_n=n\cdot\frac{c}{2L}\,.
    \label{eq:_qm_enspec}
\end{equation}
As previously stated, the $\psi_n(x)$ introduced above are eigenfunctions of the Hamiltonian operator and correspond to states with definite energies $E_n$. However, a generic quantum state does not necessarily coincide with a single eigenfunction. Rather, the set of Hamiltonian eigenfunctions forms a complete basis, so that any physically admissible wavefunction can be expressed as a linear superposition of them.

\section{From bound to free systems in Quantum Mechanics}\label{sec:VI}

Another key concept introduced during the experimental activity was the gradual disappearance of quantization as the confinement length of the system increases. We now have a mathematical model that supports the expectations drawn from the experimental analysis (Fig. \ref{img:lim}).

Starting from Eq. (\ref{eq:_qm_enspec}), the separations between consecutive energy levels of a particle in an infinite well or acoustic modes are:
\begin{equation}
    \Delta E_n \equiv E_{n+1}-E_n=\frac{h^2}{8mL^2} (2n+1)\,,\quad \Delta f_n\equiv f_{n+1}-f_n=\frac{c}{2L}\,,\quad ( n=1,2,...).
    \label{eq:spacing}
\end{equation}
In both cases, as $L\rightarrow\infty$, the spacing approaches $0$, the spectrum becomes continuous and the hallmark of quantization disappears. This result corroborates the experimental trend observed with successively longer straws and generalizes it to quantum systems.

\begin{figure}[H]
  \centering
    \includegraphics[scale=1.5]{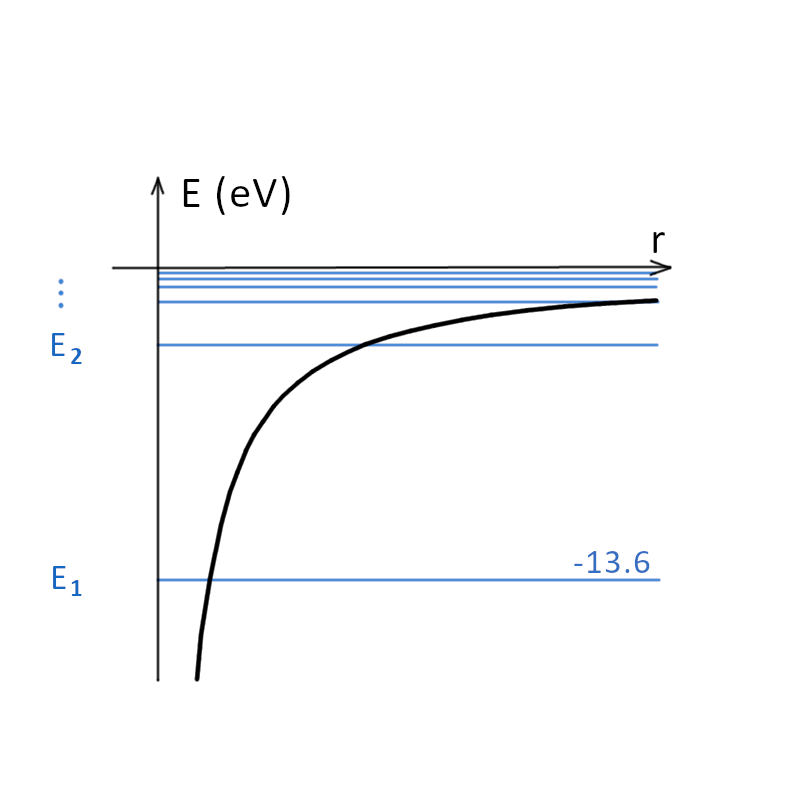}
  \caption{Energy levels of the Coulomb potential of an Hydrogen atom. As the spatial confinement decreases the energy levels become closer and closer.}
  \label{img:Hlev}
\end{figure}

To cement this idea, students can be encouraged to a qualitative examination of the Hydrogen spectrum (Fig. \ref{img:Hlev}). The energy gap between adjacent bound states diminishes with increasing principal quantum number $n$, illustrating how reduced confinement leads naturally to a quasi-continuous spectrum at the ionization threshold.

Emphazising the parallel between mathematical models and experimental results (the measured frequency spectrum of the pipe and the Hydrogen spectrum) reinforces a central message of the course: confinement produces quantization; removing confinement restores continuity. Encouraging students to map their experimental data for the drinking straw experiment onto Eq. (\ref{eq:spacing}) therefore provides a satisfying closure to this section devoted to quantization.

\section{Scope and limits of the analogy}\label{sec:VII}

The analogy between the drinking straw and the quantum infinite square well is intentionally selective. Its strength lies in highlighting a common mathematical structure: a wave confined within a finite spatial domain and subject to boundary conditions, which leads to discrete set of allowed modes. In addition, the analogy, together with the acoustic experiments, provides an immediate and intuitive way to understand how the spacing between levels decreases as the width of the well increases and confinement progressively weakens. From an educational viewpoint, this makes the origin of quantization more intuitive and experimentally accessible.

At the same time, the analogy has clear limits that should be made explicit. First, the acoustic pressure in the straw is a classical, directly measurable physical quantity, whereas the quantum wavefunction is not an observable of the same kind and, as will be shown shortly, is interpreted probabilistically. Second, the oscillating air column describes a collective classical phenomenon, while the quantum model concerns the state of a microscopic particle. 

Third, the Hydrogen atom is not literally a one-dimensional infinite square well. The latter is introduced here only as a pedagogically useful approximation that preserves the key ideas
of confinement, boundary conditions, and discrete spectra in a mathematically simple form.
In reality, Hydrogen is a three-dimensional quantum problem with central symmetry, in which
the electron is bound by the Coulomb potential of the proton. Owing to this symmetry, the
Schr\"odinger equation can be separated into a radial part and an angular part. The simplified
one-dimensional model is therefore not intended as a realistic representation of the atom, but
as an introductory step toward more adequate descriptions. This point is taken up again in the
following subsection, where a first extension beyond the infinite square well is briefly outlined.

\subsection{Limits of the analogy as a simplified model of the Hydrogen atom and possible extensions}
\label{sec:limits_extensions}

The analogy between the drinking straw and the infinite square well is intentionally minimal and pedagogically oriented. At the same time, however, the analogy should not be pushed too far. The infinite square
well is not a realistic model of the Hydrogen atom, but only a deliberately simplified introductory approximation. Its main limitation is spectral. In the one-dimensional infinite square well, the allowed energies scale as
\begin{equation}
E_n^{\mathrm{box}}=\frac{h^2}{8mL^2}n^2,
\qquad n=1,2,3,\dots
\label{eq:box_levels_limit_section}
\end{equation}
so the spacing between neighboring levels increases with $n$. By contrast, the Hydrogen atom exhibits
\begin{equation}
E_n^{\mathrm{H}}=-\frac{13.6\ \mathrm{eV}}{n^2},
\qquad n=1,2,3,\dots
\label{eq:hydrogen_levels_limit_section}
\end{equation}
so that the bound states become progressively closer and closer as the ionization threshold
$E=0$ is approached.

This difference reflects a deeper physical distinction. In the infinite square well, the particle
is confined by rigid, impenetrable boundaries inside a region of fixed size. In the Hydrogen
atom, instead, the electron is bound by a long-range Coulomb attraction, which gradually
tends to zero at large distance. As a result, highly excited states extend over increasingly
larger spatial regions, and the spectrum becomes denser near the continuum threshold.
The straw activity can therefore help students grasp the qualitative link between weaker
confinement and smaller level spacing, but it cannot reproduce the actual spectral structure
of Hydrogen.

A natural extension of the present teaching sequence is therefore to move from the infinite
square well to a simple finite radial model. Even without entering into the full mathematical derivation, it is useful to note that a piecewise-constant radial potential already captures some features that the infinite square well entirely misses: a finite ionization threshold, exponentially decaying tails of the eigenfunctions outside the classically allowed region, and a first qualitative tendency toward level crowding near $E=0$.

In a preliminary extension of the present modelling strategy, we explored a simple
two-step radial potential, characterized by an inner region of depth $V_1$ and radius $R_1$,
followed by a broader and shallower region of depth $V_2$ extending up to $R_2$. In one
representative case, the parameters were of the order of
\[
V_1 \approx 27.4\ \mathrm{eV},\qquad
V_2 \approx 3.4\ \mathrm{eV},\qquad
R_1 \approx 2.25\,a_0,\qquad
R_2 \approx 17.5\,a_0,
\]
where $a_0$ is the Bohr radius. These values should be regarded only as an exploratory proof of concept, introduced to show how a finite radial piecewise-constant model can qualitatively improve the
description of the Hydrogen spectrum.

\begin{figure}[H]
\centering
\includegraphics[width=0.88\textwidth]{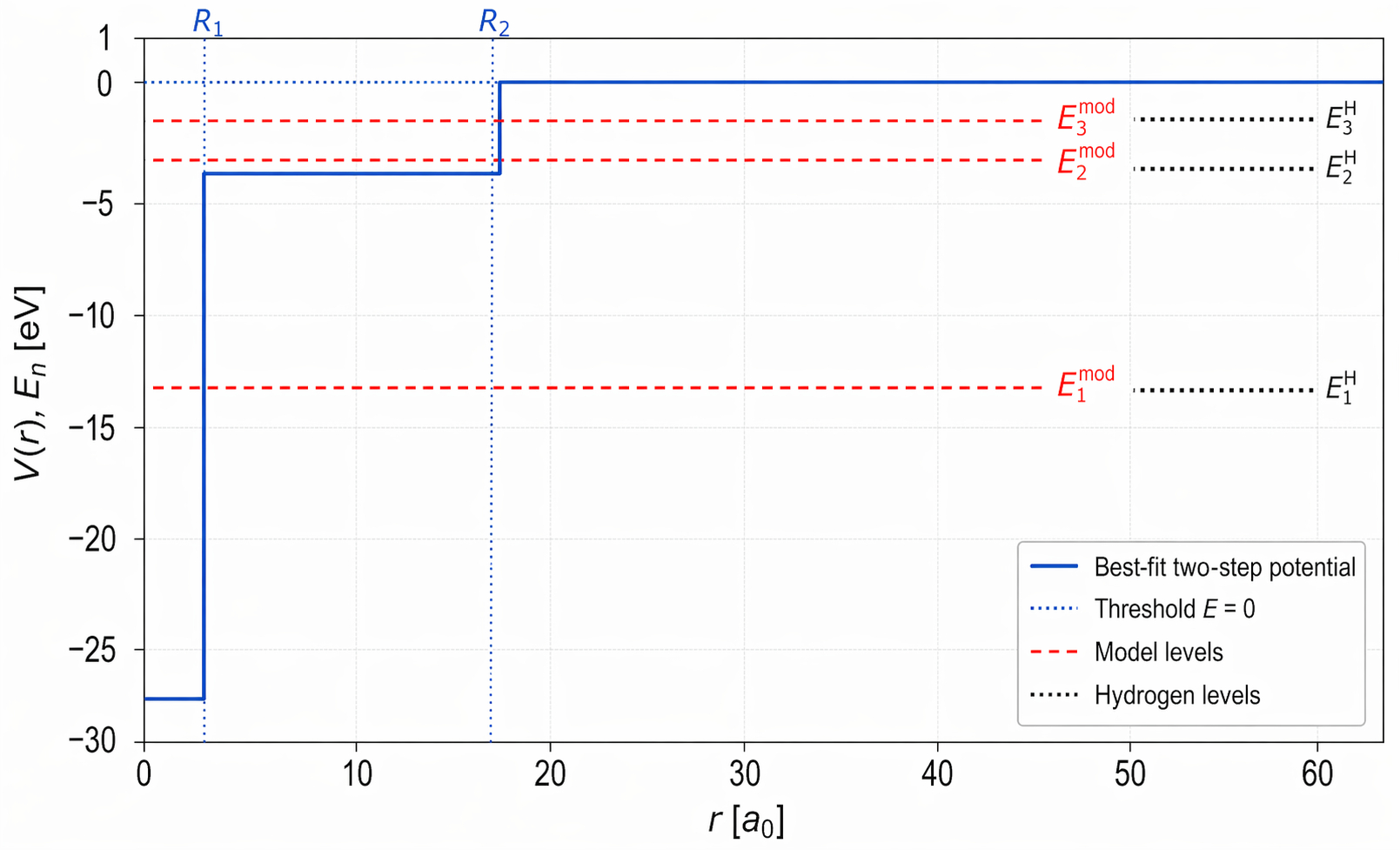}
\caption{Example of a simple radial two-step potential used as a qualitative extension of the
infinite-well model. The inner and outer radial scales are
$R_1 \approx 2.25\,a_0$ and $R_2 \approx 17.5\,a_0$, while the corresponding effective depths are
$V_1 \approx 27.4$ eV and $V_2 \approx 3.4$ eV. The first three energy levels of the model are compared with the corresponding Hydrogen reference levels. }
\label{fig:twostep_radial_model}
\end{figure}

An example of such an extension is shown in Fig.~\ref{fig:twostep_radial_model}, where the
Coulomb interaction is replaced by a two-step radial potential consisting of a deeper inner region and a broader, shallower outer region.

Exploratory calculations suggest that such a simple radial model can already reproduce the first Hydrogen levels at a qualitatively reasonable level. In our case, the first three energies were found to be approximately
\[
E_1^{\mathrm{mod}}\approx -13.5\ \mathrm{eV},\qquad
E_2^{\mathrm{mod}}\approx -3.0\ \mathrm{eV},\qquad
E_3^{\mathrm{mod}}\approx -1.6\ \mathrm{eV},
\]
to be compared with the Hydrogen values
\[
E_1^{\mathrm{H}}=-13.6\ \mathrm{eV},\qquad
E_2^{\mathrm{H}}=-3.4\ \mathrm{eV},\qquad
E_3^{\mathrm{H}}=-1.51\ \mathrm{eV}.
\]
These results are only meant as an initial validation. Their educational value lies in showing that the main feature missing from the infinite square well is not related to its depth, which, as our case shows, can also be taken to be finite, but rather to the absence of an extended outer attractive region of the potential. In the present model the outer region extends up to $\sim 18$ Bohr radii. In other words, the overall profile of the energy spectrum seems to depend primarily on the more extended outer region of the potential, rather than on the detailed structure or depth of its inner part.

Within this perspective, the infinite square well retains a precise and important role: it is an excellent first model for introducing the fact that confinement of a wave leads to quantization. Its limitations are not a weakness of the teaching sequence; on the contrary, they provide an
opportunity to discuss how scientific models are refined. The comparison between the straw, the infinite square well, and a first finite radial model can help students appreciate that physical models are useful not because they are exact, but because they isolate relevant features of a phenomenon at different levels of complexity.

A more detailed mathematical treatment of this extended radial modelling approach, including the associated radial eigenfunctions and a systematic discussion of its didactical implications, will be developed in a subsequent paper.

\section{Introducing the wavefunction interpretation}\label{sec:VIII}
Our analogy also provides an opportunity to introduce wavefunctions in quantum mechanics and to spark a discussion about their interpretation.

Since every pressure mode they measured in the straw has a one-to-one twin among the eigenfunctions of an infinite square well, students can visualize the pure-state wavefunctions of the quantum problem (see section \ref{sec:II} and Fig. \ref{img:mod}). Having drawn that visual bridge, the natural following step is to pose the question: what is the physical meaning of the wavefunction?

Whereas the physical interpretation of the pressure modes in the pipe is relatively intuitive, establishing a clear connection between the wavefunction and a quantum particle proves more challenging from both a scientific and pedagogical perspective. At this stage, the most widely accepted interpretation of the wavefunction (the Copenhagen interpretation) can be introduced: it is a mathematical tool whose squared modulus defines the probability of finding the particle within a given spatial interval.
\begin{equation}
dP = \lvert \Psi \rvert ^2  dx
    \label{eq:copen}
\end{equation}
\begin{figure}[H]
  \centering
    \includegraphics[scale=0.3]{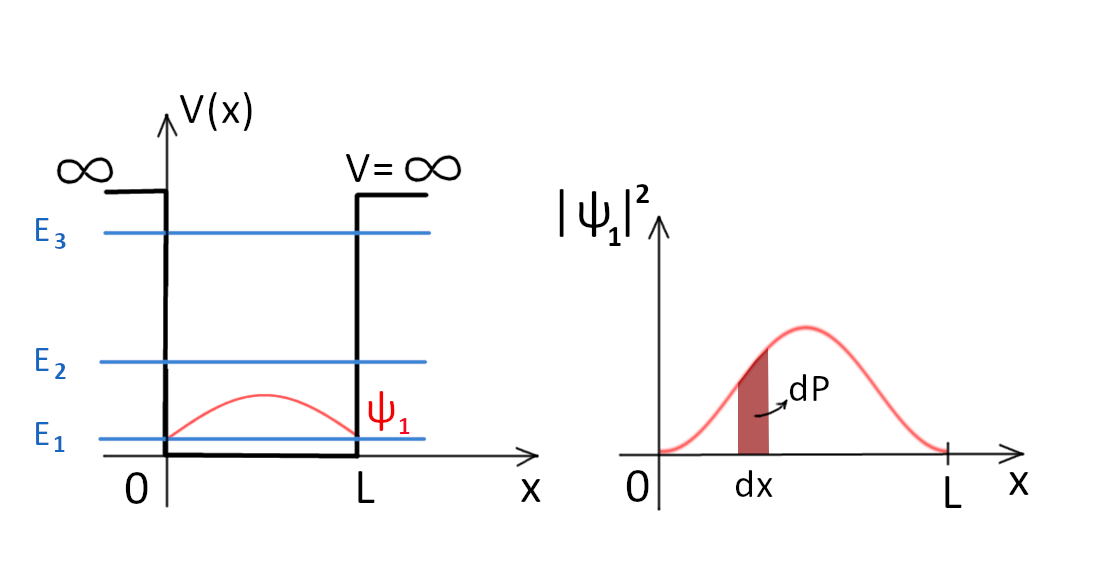}
  \caption{The probabilistic interpretation of the wavefunction.}
  \label{img:psi2}
\end{figure}
The ground-state eigenfunction, $\Psi_1$, can be used to provide a visual representation (Fig. \ref{img:psi2}) of the probabilistic interpretation expressed in Eq. (\ref{eq:copen}).   This can be followed by a classroom discussion that encourages students to think critically. They may be invited to explore conceptual tensions or open questions \mycite{bib1} that arise from the probabilistic nature of the interpretation.

Such an educational approach is particularly valuable because it fosters a deeper understanding of the nature of science itself and of how scientific knowledge evolves. Encouraging critical thinking is essential for cultivating scientific literacy. In contrast, presenting a theory as a finished and unquestionable construct risks conveying the misleading idea that science is a collection of absolute truths, rather than a dynamic process grounded in questioning, refinement, and continuous exploration.

An example of a provocative issue that might arise from the Copenhagen interpretation of the wavefunction could be:  in a bound atomic state the electron is described by a spatially extended wavefunction whose squared modulus yields a non-zero probability density over a range of positions. Interpreted naively, this would seem to imply that the electron’s charge occupies different locations in a finite region at successive instants and is therefore subject to acceleration. Classical electrodynamics predicts that any accelerating charge must emit electromagnetic radiation and lose energy, yet experimentally atoms in stationary states remain stable and non-radiative.

\section{Heisenberg Uncertainty relations}\label{sec:IX}
The Heisenberg uncertainty relations are implicitly hidden into the results we have just derived. A clear indication is that the ground-state energy $E_1$ is not zero, something classical mechanics would never predict. Classically, a particle could sit perfectly still inside the well and therefore have zero energy. In quantum mechanics this situation is forbidden by the uncertainty relation
\begin{equation}
    \sigma_x\,\sigma_p\sim\,\hbar/2\, ,
    \label{eq:heis}
\end{equation}
where $\sigma_x$ and $\sigma_p$ are the standard deviations of the position and momentum wavefunctions \mycite{bib6}. If the particle were truly at rest, its momentum would be sharply defined, driving $\sigma_p\rightarrow 0$. Eq. (\ref{eq:heis}) would then force $\sigma_x\rightarrow\infty$, which contradicts the finite spatial spread clearly visible in the ground-state wavefunction $\Psi_1(x)$ (Fig. \ref{img:boxlev}).
The non zero “zero-point” energy is therefore a direct, tangible manifestation of the Heisenberg relation.

 A simple numerical estimate makes this point more concrete. Consider an electron confined in a one-dimensional infinite well of width $L = 1.0 \times 10^{-10}\,\mathrm{m}$, a length scale comparable to atomic dimensions. Using Eq. (\ref{eq:_qm_enspec}), the ground-state energy is

\begin{equation}
    E_1 = \frac{h^2}{8mL^2} \approx 6.0 \times 10^{-18}\,\mathrm{J} \approx 37.6\,\mathrm{eV}.
\end{equation}

This value is clearly different from zero and shows that spatial confinement alone implies a finite minimum energy. In the infinite square-well model, the ground-state energy of $37.6 \,\mathrm{eV}$ is entirely kinetic. Interestingly, this value is of the same order of magnitude as the mean kinetic energy of the electron in the ground state of the Hydrogen atom, $13.6 \,\mathrm{eV}$. The latter can be inferred either from the Coulomb potential evaluated at the Bohr radius together with the total ground-state energy, or more rigorously from the virial theorem.

\section{Results}\label{sec:X}
The ``Quantum Straw" sequence was carried out in three independent five-hour courses that involved nearly 100 upper-secondary students working in 32 small teams. The discussion below summarizes two complementary data sets: the quantitative accuracy of the experimental work and the qualitative evaluation obtained through 52 anonymous student comments collected at the end of the courses.

\subsection{Experimental accuracy}
Students measured the first five resonance frequencies of a 19.9 cm drinking-straw and compared them with theoretical predictions. The averaged relative deviation was below 3\% for every mode (Table \ref{tab:fr}). These values confirm that low-cost, smartphone-based spectroscopy provides results that are fully consistent with acoustics. The ability to obtain such agreement after only a brief introduction indicates that the experimental protocol is both robust and readily transferable to the educational setting. 

\subsection{Student perception of the learning experience}

To examine students’ perceptions of the learning experience, we analyzed 52 anonymous written comments collected at the end of the three courses. The analysis was conducted as an exploratory descriptive thematic coding rather than as a fully formal qualitative study. First, comments that contained no evaluative content (e.g., blank entries or isolated punctuation marks) were excluded. The remaining comments were read repeatedly and segmented into meaning units expressing evaluative judgments about the course. In a second step, these units were grouped into recurrent thematic categories. The final categories reported in Table II were: engagement/interest, lecturer effectiveness, conceptual gain, and difficulty/pacing issues. Because a single comment could contain more than one evaluative aspect, comments were allowed to contribute to more than one category. The coding was carried out by the author, who also taught the courses. Therefore, no inter-rater reliability coefficient was computed, and the qualitative findings should be interpreted as exploratory indications of recurring student perceptions rather than as a fully independent qualitative validation of the intervention.

\begin{figure}[H]
  \centering
    \includegraphics[scale=0.49]{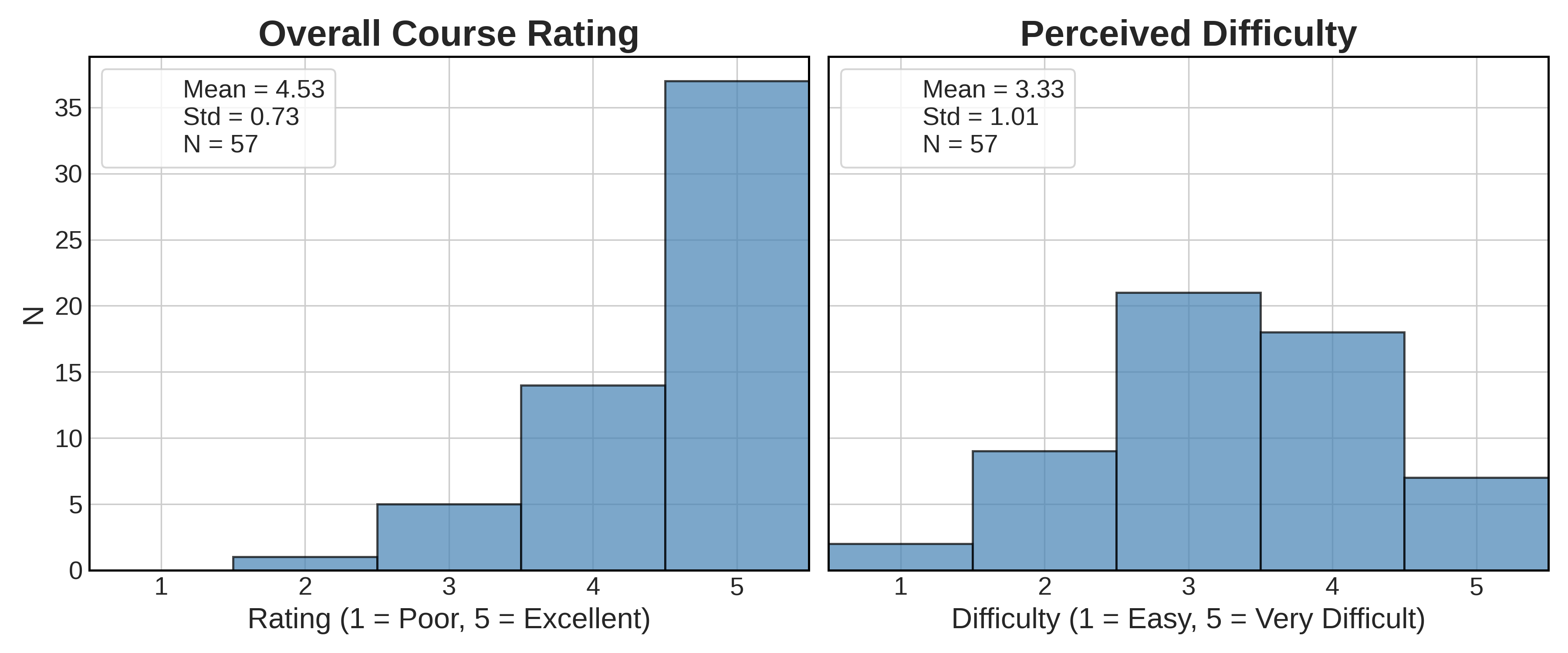}
  \caption{The course rating and the perceived difficulty as reported by 57 students.}
  \label{img:feed}
\end{figure}

\begin{table}[H]
\centering
\caption{}
\begin{ruledtabular}
\begin{tabular}{ c c c}
 \textbf{Theme} & \textbf{Typical adjectives / verbs used by students} & \textbf{Frequency}  \\ \hline
Engagement  & interesting, engaging, fantastic & 41/52 (79\%) \\
Lecturer effectiveness & clear, brilliant, supportive & 37/52 (71\%) \\
Conceptual gain & understand, deepen, eye-opening & 29/52 (56\%)\\
Difficulty and pacing issues & difficult, too fast, math taken for granted & 4/52 (8\%)
\end{tabular}
\end{ruledtabular}
\label{tab:stud}
\end{table}
Overall sentiment was strongly positive: 35 comments contained explicitly positive evaluations with no reservations, 13 combined appreciation with concerns about mathematical prerequisites or time constraints, three were neutral, and one was purely negative (a single ``too fast”). Students particularly valued the analogy-based approach and the opportunity to link equations to hands-on evidence\mycite{bib4}. 
The course received a high overall rating, averaging 4.5 on a 5-point Likert scale, while students judged its difficulty to be moderate, with a mean score of 3.3 (Fig. \ref{img:feed}).
\subsection{Learning outcomes}
Taken together, the quantitative and qualitative results suggest that the strategy meets its two primary objectives.
\begin{enumerate}
    \item Conceptual understanding. The majority of students reported a clearer grasp of why confinement leads to quantization, corroborated by their successful prediction of how resonance spacing changes with straw length.
    \item Affective impact. High levels of engagement and lecturer appreciation indicate that the activity fosters a positive attitude toward quantum physics, an outcome that is often difficult to achieve at this level.
\end{enumerate}

\subsection{Areas for refinement}
The feedback suggests two practical adjustments. First, it may be beneficial to allocate additional time to unpack the mathematical formalism, particularly for students with less prior preparation. Additionally, offering an optional slower-paced session or a pre-course refresher on sinusoidal functions could help address concerns related to the pacing of the material.

\section{Conclusions}\label{sec:XI}
This work has presented and tested an analogy-based approach for introducing fundamental quantum concepts. By leveraging the formal and conceptual parallels between sound waves in a straw and the quantum behavior of particles in a potential well, the activity enables students to explore energy quantization through direct experimentation and intuitive reasoning. The integration of low-cost tools and accessible mathematical models contributes to both the feasibility and pedagogical value of the method.

The results suggest that the analogy not only supports the understanding of abstract quantum ideas, such as wavefunction and energy levels, but also promotes student engagement and curiosity. The experimental accuracy obtained, combined with strongly positive student feedback, indicates that this approach can successfully bridge the gap between hands-on learning and theoretical insight.

Furthermore, the activity opens opportunities to introduce deeper scientific concepts, such as the Copenhagen interpretation, the Heisenberg uncertainty principle and the role of boundary conditions in wave phenomena, within a meaningful and motivating context.

Future implementations may benefit from adjustments in pacing and mathematical scaffolding to better support diverse student backgrounds. Nonetheless, the overall findings reinforce the potential of well-designed analogies as powerful tools in science education, capable of making complex topics more accessible and intellectually rewarding.

\begin{acknowledgments}
We gratefully acknowledge the ``Scuola di Formazione Luigi Lagrange" for the opportunity to include this topic among its course offerings, which enabled us to test and refine the proposed analogy-based approach.
\end{acknowledgments}

\section*{Data Availability}
The entire dataset supporting the results of this study is published in the article.


\makeatletter
\let\auto@bib\@empty
\makeatother

\end{document}